# Preferences for COVID-19 vaccine: Evidence from India[1]

Prateek Bansal[*], Alok Raj, Dhirendra Mani Shukla, Naveen Sunder

[1]Department of Civil and Environmental Engineering, National University of Singapore, Singapore
[2]Production, Operations & Decision Sciences, XLRI Xavier School of Management, Jamshedpur, India
[3]Business Policy and Strategy, Indian Institute of Management, Udaipur, India
[4]Department of Economics, Bentley University, Waltham, Massachusetts, USA

*corresponding author (pb422@cornell.edu)

**Abstract**

India's mass vaccination efforts have been slow due to high levels of vaccine hesitancy. This study uses data from an online discrete choice experiment with 1371 respondents to rigorously examine the factors shaping vaccine preference in the country. We find that vaccine efficacy, presence of side effects, protection duration, distance to vaccination centre and vaccination rates within social network play a critical role in determining vaccine demand. We apply a non-parametric model to uncover heterogeneity in the effects of these factors. We derive two novel insights from this analysis. First, even though, on average, domestically developed vaccines are preferred, around 30 percent of the sample favours foreign-developed vaccines. Second, vaccine preference of around 15 percent of the sample is highly sensitive to the presence of side effects and vaccination uptake among their peer group. These results provide insights for the ongoing policy debate around vaccine adoption in India.

---

[1] This work was supported by the Xavier School of Management, Jamshedpur (India) - Grant number - XLRG2013.

# 1. Introduction

India, the context that we investigate in the current study, is the country with the second-highest incidence of COVID-19. As of early August 2021, the country has recorded more than 30 million cases and over 400,000 deaths (Worldometers, 2021). The Indian government has tried to prevent the spread of the virus through measures such as mask mandates, lockdowns, and mass vaccination. Even though mass vaccination efforts hold the highest potential for bringing an end to the current pandemic (Schwarzinger et al., 2021), vaccination rates remain low in India. As of now, less than 8 percent of all Indians have been fully vaccinated (Bloomberg, 2021). Vaccine hesitancy, estimated to be between 29 and 42 percent (Lazarus et al., 2021), is the primary driver of low vaccination rates in India. There is an urgent need to understand characteristics shaping preferences for the COVID-19 vaccine in the Indian context.

Our analysis is one of the first and most detailed investigations of the factors affecting COVID-19 vaccine preferences in India. Using a Discrete Choice Experiment (DCE), we quantify the sensitivity of consumers' vaccine preference relative to changes in various attributes such as efficacy, protection duration, side effects, price and administration location. Finally, we explore heterogeneities in the effectiveness of drivers of vaccine preferences using a non-parametric empirical model. Such a demand-side analysis is timely and critical because countries need to ramp up vaccination efforts in the face of the emergence of new variants. The findings of our study have the potential to shape ongoing policy discussions in India and other developing countries.

# 2. Experiment design and data collection

**a. Sample and data collection**

The data for this study comes from an online discrete choice experiment (DCE) conducted between May and June 2021. The sample consists of individuals who are i) above 18 years of age, ii) have not yet received COVID-19 vaccine, and iii) residing in one of the following five states – Maharashtra, Tamil Nadu, Uttar Pradesh and West Bengal[2]. We use a quota sampling approach, where respondents were stratified based on gender, marital status, and age. The final analysis sample consists of 1371 observations. The summary statistics are presented in Table 1 - male, single, and younger age groups are slightly overrepresented, but our sample is adequately representative of the Indian population.

---
[2] The respondents come from a panel enlisted by MarketXcel, a professional market research agency in India.

Table 1. Demographic and spatial distribution across sample and population (N=1371).

| Attributes | Sample | 2011 Indian census | State (Zone) | Sample | Rank in confirmed COVID cases (June 14, 2021) |
|---|---|---|---|---|---|
| *Age* | | | Maharashtra (West) | 23% | 1 |
| 18 to 39 year | 69% | 57% | Tamil Nadu (South) | 23% | 4 |
| 40 to 60 year | 23% | 32% | Uttar Pradesh (North) | 15% | 6 |
| Above 60 year | 8% | 11% | West Bengal (East) | 19% | 7 |
| *Marital Status* | | | Delhi (North) | 20% | 8 |
| Single | 32% | 21% | | | |
| Married | 66% | 78% | | | |
| Other | 2% | 1% | | | |
| *Gender* | | | | | |
| Male | 59% | 51% | | | |
| Female | 41% | 49% | | | |

### b. Experiment design

In this study, we design a DCE to study the factors affecting the preferences for the COVID-19 vaccine. The selection of attributes is guided by a literature review of journal articles (Borriello et al., 2021; Dong et al., 2020; McPhedran and Toombs, 2021). The levels of the attributes are chosen such that i) they span the entire attribute support, and ii) vaccine profiles are comparable to the ones that are available in India. Table 2 presents details on the attributes included in our experiment.

Table 2. Attribute levels in the DCE.

| Attributes | Levels |
|---|---|
| Effectiveness of vaccine | 80% <br> More than 90% |
| Developer | Domestic <br> Imported |
| Out-of-pocket cost | INR 0 <br> INR 400 (US$5.4) <br> INR 800 (US$10.8) <br> INR 1200 (US$16.2) |
| Side effects | No side effect <br> Fever or headache |
| Duration of protection | 6 months <br> 12 months <br> 18 months <br> 24 months |
| Place of vaccine administration | At a government hospital <br> At a private hospital <br> At your home |
| The proportion of friends and family members who has taken the vaccine | 10% <br> 50% <br> 90% |

In the experiment, respondents are asked to choose their preferred vaccine between two vaccine alternatives. We generated a DCE design using the D-efficient approach with zero priors (Rose and Bliemer, 2009), which leads to efficient estimation by extracting maximum information from the data. Our experiment consists of six blocks with six choice scenarios per block, and each respondent was shown a randomly selected block[3]. An example choice scenario is presented in Table 3.

Table 3. An example of the choice situation presented to respondents.

|  | Vaccine 1 | Vaccine 2 |
|---|---|---|
| Effectiveness of vaccine | 80% | More than 90% |
| Vaccine developer | Domestic | Imported |
| Purchase price | ₹ 1200 | ₹ 800 |
| Side effects | Fever or headache | No side effect |
| Duration of protection | 12 months | 6 months |
| Place of vaccine administration | Home | Private hospital |
| The proportion of friends and family members who has taken the vaccine | 10% | 90% |

## 3. Empirical strategy

We use a conditional logit (CL) model to estimate the effect of attributes on the preference for COVID-19 vaccine. The indirect utility under this specification can be expressed as:

$$U_{itj} = x'_{itj}\boldsymbol{\delta} + \varepsilon_{itj}. \qquad (1)$$

In Equation (1), $U_{itj}$ is the indirect utility of individual $i$ from choosing alternative $j$ in choice scenario $t$, $\boldsymbol{\delta}$ is the vector of marginal utilities, $x_{itj}$ is the corresponding attribute vector (presented in table 3), and $\varepsilon_{itj}$ is idiosyncratic error term with standard Gumbel distribution. Thus, individual $i$'s probability of choosing alternative $j$ in choice scenario $t$ is:

$$Pr_{itj} = \frac{\exp(U_{itj})}{\sum_{\forall k}\exp(U_{itk})} \qquad (2)$$

Additionally, we quantify the unobserved heterogeneity in main effects across different parts of the population using a non-parametric logit mixed logit (LML) model (Bansal and Daziano, 2018; Train, 2016). The indirect utility in the LML specification is:

$$U_{itj} = x'_{itj}\boldsymbol{\delta} + w^{R'}_{itj}\boldsymbol{\beta}^R_i + \varepsilon_{itj}. \qquad (3)$$

A key point of departure from the CL model is that the LML specification contains random parameters ($\boldsymbol{\beta}^R_i$), which are assumed to have a discrete mixing distribution over their finite support set $\boldsymbol{\Omega}$ (or multi-dimensional grid). The joint probability mass function of random parameters in LML is specified as follows:

---

[3] There is no consensus in the literature against including or excluding the opt-out alternative (Ryan and Skåtun, 2004). To circumvent potential misinterpretation of the opt-out alternative and avoid modelling challenges arising from a zero-level alternative, we do not present it to respondents. If we were interested in welfare estimation, an opt-out alternative could have been included.

$$\Pr(\boldsymbol{\beta}_i^R = \boldsymbol{\beta}_r^R) = \frac{\exp\left[z(\boldsymbol{\beta}_r^R)'\boldsymbol{\alpha}\right]}{\sum_{s\in\Omega}\exp\left[z(\boldsymbol{\beta}_s^R)'\boldsymbol{\alpha}\right]} \qquad (4)$$

In Equation (4), $z(\boldsymbol{\beta}_r^R)$ is assumed to be a spline function, and $\boldsymbol{\alpha}$ is the corresponding vector of parameters. The LML model is estimated using maximum simulated likelihood estimator, which is described in Bansal et al. (2018). Standard errors are obtained using bootstrapping.

## 4. Results

The results from the CL model (columns 1-2, Table 4) indicate that vaccine efficacy and side effects are critical determinants of preferences for COVID-19 vaccine. The odds of accepting a vaccine with 90 percent or higher effectiveness is 1.26 times higher than one with 80 percent effectiveness. Individuals also have a higher likelihood of choosing a vaccine with no side effects (odds ratio = 1.35), and a longer duration of protective effects (odds ratio = 1.16). We show that Indian consumers have a lower probability of picking a vaccine developed outside of India (odds ratio = 0.93), while their odds of getting the vaccine when all of their peer group has been fully vaccinated is, on average, 1.41 times higher than those with an entirely unvaccinated peer group.

We further probe for heterogeneities in the observed results using a non-parametric LML model. Instead of a single coefficient (CL estimates), the LML model provides a probability (cumulative) distribution function of the odds ratios – the results for each attribute are presented in Figure 1. The graphs in Figure 1 demonstrate that the distribution of odds ratio varies significantly from the CL estimate, although the mean odds ratio from the LML model are similar in magnitude to the CL estimates (see columns 2 and 5 of Table 1). The heterogeneity in effects, as visible in Figure 1, are statistically confirmed by the significant estimates of standard deviations (column 4 in Table 4).

We uncover some important patterns from the LML estimates. Although the coefficient estimates of the CL model (column 1 of Table 4) suggests that there is a preference for domestically developed vaccines and vaccination at home, the cumulative distribution function from LML (Figure 1), however, indicates that 31.2 and 29.7 percent of individuals preferred foreign-developed vaccines and vaccination in private/government hospitals, respectively. The LML results also demonstrate that around 12 percent of the sample have odds ratio of the vaccinated social network above five, while around 17 percent have a strong preference for a vaccine with no side effects (odds ratios > 6), which are very different from the CL estimates.

Table 4. Results of conditional logit and logit mixed logit model (N=1371).

| Attributes | Conditional Logit | | Logit Mixed Logit | | |
|---|---|---|---|---|---|
| | (1) Coeff (Std. Err.) | (2) Odds ratio | (3) Mean of Coeff. (Std. Err.) | (4) Std. Dev. Of Coeff. (Std. Err.) | (5) Mean odds ratio |
| Alternative-specific constant | -0.322*** (0.026) | | -0.353*** (0.031) | | |
| 90% or more effective (base: 80%) | 0.231*** (0.024) | 1.26 | 0.338*** (0.031) | 0.237*** (0.030) | 1.40 |
| Foreign developer (base: domestic) | -0.072*** (0.024) | 0.93 | -0.067* (0.041) | 0.225*** (0.028) | 0.94 |
| Out of pocket cost (unit increase: INR 100) | -0.051*** (0.003) | 0.95 | -0.064*** (0.005) | | 0.94 |
| No side effect (base: fever/headache) | 0.301*** (0.024) | 1.35 | 0.439*** (0.029) | 0.687*** (0.040) | 1.55 |
| Protection duration (unit increase: 6 month) | 0.148*** (0.014) | 1.16 | 0.243*** (0.019) | 0.306*** (0.036) | 1.27 |
| Vaccine administration at home (base: hospital) | 0.089*** (0.029) | 1.09 | 0.073* (0.049) | 0.349*** (0.053) | 1.08 |
| Vaccinated friends/family (base:0, unit increase: all) | 0.341*** (0.044) | 1.41 | 0.491*** (0.052) | 0.537*** (0.042) | 1.63 |
| Loglikelihood | -5171.5 | | -5013.6 | | |

***p-value < 0.01, **0.01 < p-value < 0.05, *0.05 < p-value < 0.15.

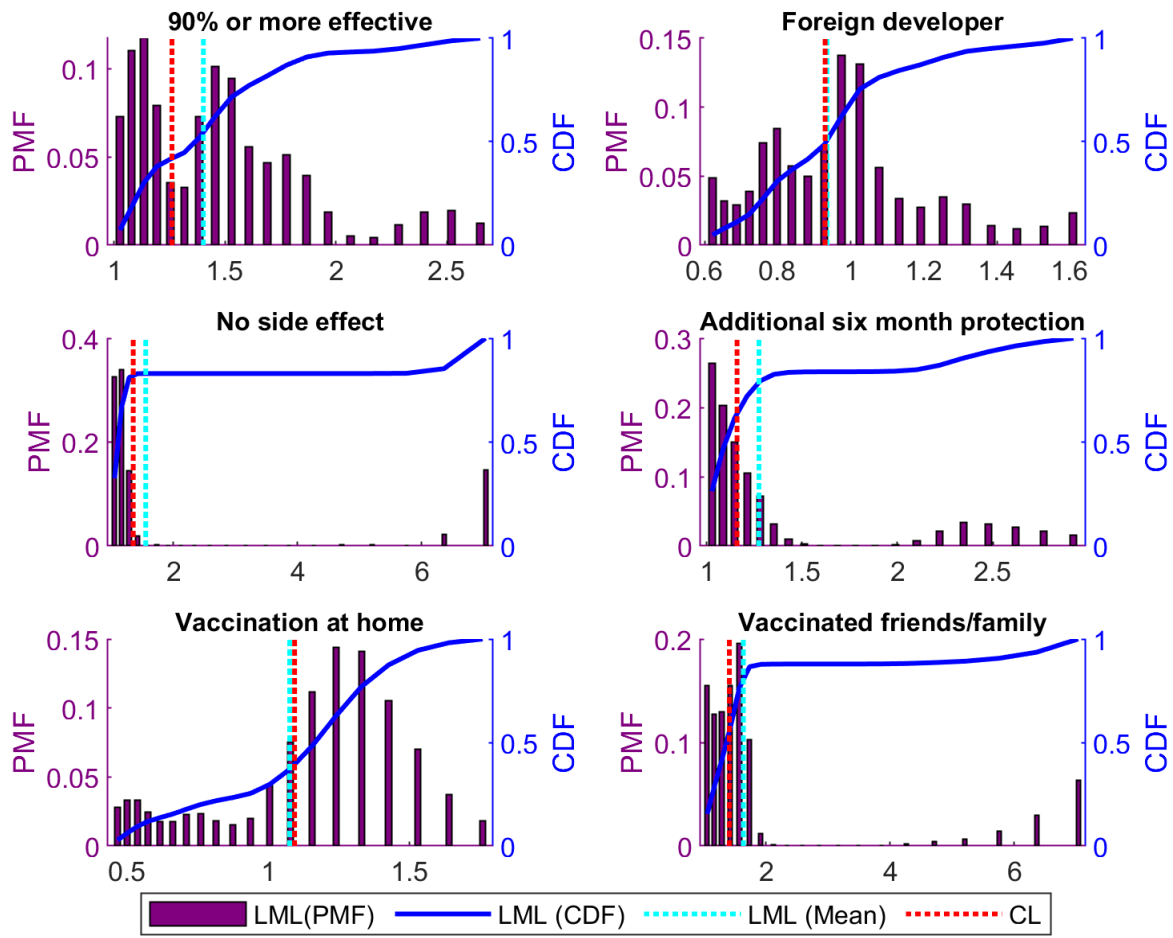

**Figure 1.** The estimated distribution of odds ratios (LML: logit mixed logit, CL: conditional logit; PMF: probability mass function; CDF: cumulative distribution function)

# 5. Discussion and Conclusions

The results from this study are largely congruent with recent evidence from other contexts - vaccine efficacy, ease of inoculation, social networks and presence of side effects are important determinants of preferences for COVID-19 vaccine (see Dong et al., 2020; Schwarzinger et al., 2021). An encouraging result of our analysis is that individuals are much more likely to choose a vaccine with more than 90 percent efficacy - two out of the three vaccines approved for use in India meet this criterion.

Although the aggregate effects favour a domestically developed vaccine, we also demonstrate that there is a sizable part of the population (31.2 percent) that has a higher likelihood of selecting a foreign-developed vaccine. This suggests that giving individuals a choice regarding vaccine type might significantly increase their uptake. This is important from a policy perspective since at the moment most facilities in India do not provide such a choice.

Our findings also suggest that providing vaccines in people's homes might further increase uptake. This insight suggests that the government should plausibly increase door-to-door vaccination efforts, especially in locations with low health facility density. It is important that these efforts should complement, and not replace, the current strategy of vaccination at health centres – this is because our data suggest that a considerable proportion (30 percent) of the population still prefer to get vaccinated at hospitals.

Our results demonstrate that large sections of the population are concerned about vaccine side effects (17 percent) and are influenced by vaccine uptake among their peer group (12 percent). These sub-populations are critical because to reach herd immunity the Indian government has to tailor policies to cater to these potential *late adopters*. Examples of such policies include subsidized (or free) care to address any issues related to vaccine side effects, and effective communication strategies to demonstrate that the benefits of taking a vaccine significantly outweigh the risks.

The present study provides timely quantitative evidence on factors that shape vaccine preferences in India, one of the nations that is in dire need of ramping up vaccination efforts. It is also one of the first detailed studies of determinants of preferences for COVID-19 vaccines in the South Asian context and will contribute to policy discussions on ways to expedite COVID-19 vaccine delivery in the region.